\documentclass[a4paper,fleqn]{cas-sc}

\usepackage[numbers]{natbib}

\def\tsc#1{\csdef{#1}{\textsc{\lowercase{#1}}\xspace}}
\tsc{WGM}
\tsc{QE}
\tsc{EP}
\tsc{PMS}
\tsc{BEC}
\tsc{DE}

\begin{document}

\let\WriteBookmarks\relax
\def\floatpagepagefraction{1}
\def\textpagefraction{.001}

\shorttitle{Comment on Zhou and Yu (2026)}
\shortauthors{A. Pekcan}

\title [mode=title]
{Comment on "Space-time shifted solitons and solution interactions
for a generalized nonlocal nonlinear Schr\"odinger equation"}

\author[1]{Asl{\i} Pekcan}[orcid=0000-0003-3712-5248]
\cormark[1]

\ead{aslipekcan@hacettepe.edu.tr}

\credit{Conceptualization, Methodology, Formal analysis, Investigation,
Writing-original draft, Writing-review \& editing}

\affiliation[1]{
    organization={Department of Mathematics, Faculty of Science,
    Hacettepe University},
    city={Ankara},
    postcode={06800},
    country={T\"urkiye}
}

\cortext[cor1]{Corresponding author}

\maketitle

\begin{abstract}
Zhou and Yu [Appl. Math. Lett. 178 (2026) 109937] studied the
space-time shifted NLS equation
$iq_t(x,t)=q_{xx}(x,t)-2\sigma q^2(x,t)
 \bar{q}(x_0-x,t_0+\lambda t),\,\, \lambda=\pm 1$. They claimed that for both choices of $\lambda$ the equation is integrable.
For $\lambda=-1$, however, the equation is not compatible with the standard NLS system and hence is not an integrable reduction of that system.
The Hirota expansion used in the commented article does not truncate as stated; consequently
 the given one-soliton solution is not a solution for $\lambda=-1$.
For $\lambda=1$, consistency of the shifted reduction of NLS system requires $t_0=0$ as a
mandatory condition. This condition is not reported; instead $t_0$ is incorrectly treated as an arbitrary constant in the paper.
The notation used in the Hirota form is misleading,
and we provide a corrected form. Additionally, the equivalence between nonlocal equations with real shifts and their unshifted counterparts under translation of independent variables is clarified.
\end{abstract}

\begin{highlights}
    \item The equation with $\lambda = -1$ is not a consistent reduction of the standard integrable NLS system.
   \item For $\lambda = 1$, reduction consistency requires the omitted condition $t_0= 0$.
    \item The Hirota expansion for the equation with $\lambda = -1$ fails at order $\varepsilon^3$.
    \item The presented one-soliton solution formula does not solve the equation with $\lambda = -1$.
    \item Shifted-conjugate profiles must preserve localization and periodicity.
\end{highlights}


\section{The equation under consideration}

The equation studied in Ref.~\cite{ZhouYu2026} is
\begin{equation}
 iq_t=q_{xx}-2\sigma q^2\bar{q}(x_0-x,t_0+\lambda t),
 \qquad \lambda=\pm 1,                         \label{commented}
\end{equation}
where $x_0,t_0\in\mathbb{R}$.  In that paper, this equation is described as an integrable
space-time shifted nonlocal nonlinear Schr\"odinger (NLS) equation, and one- and two-soliton solutions are derived
by Hirota direct method. However, the following issues arise:
\begin{enumerate}
\item[i)] For $\lambda=-1$, the reverse space-time shifted nonlocal reduction is inconsistent with the standard integrable NLS equation. Thus, Eq.~\eqref{commented} is not an integrable reduction of the NLS equation.

\item[ii)] When $\lambda=1$, the condition $t_0=0$ becomes mandatory for compatibility with the integrable NLS system. However, Ref.~\cite{ZhouYu2026} incorrectly treats $t_0$ as an arbitrary constant.

\item[iii)] The notation used in the Hirota form is confusing since the shifted variables for the complex conjugate functions are not represented.

\item[iv)] For $\lambda=-1$, the Hirota perturbation does not truncate as claimed. In particular, the coefficient of $\varepsilon^3$ does not vanish for arbitrary wave number $k$. Hence, the given one-soliton expression is not a solution for Eq.~\eqref{commented} with $\lambda=-1$.

\item[v)] The shifted complex conjugate $\bar q(x_0-x,t_0+\lambda t)$ is uniquely defined in terms of $q(x,t)$ and cannot be considered as an independent second solution. Thus, the localization and periodicity features of the associated intensity patterns should be revised. It is not possible to have a bright soliton for $|q(x,t)|$ while a periodic solution for $|\bar q(x_0-x,t_0+\lambda t)|$ as given in Figs.~1(a) and 1(b).

\item[vi)] Although the constant $\sigma$ is defined as $\sigma=\pm 1$, the numerical profiles presented in Ref.~\cite{ZhouYu2026} consider values like $\sigma=4$ and $\sigma=0.1$, among others.
\end{enumerate}

\section{Consistency of the shifted nonlocal reductions}

Consider the standard integrable NLS system \cite{AKNS1974}
\begin{align}
 iq_t(x,t)&=q_{xx}(x,t)-2\sigma q^2(x,t)r(x,t),                                  \label{NLSq}\\
 ir_t(x,t)&=-r_{xx}(x,t)+2\sigma r^2(x,t)q(x,t),                                 \label{NLSr}
\end{align}
with $\sigma=\pm 1$.

We first consider the complex conjugate shifted nonlocal reduction
\begin{equation}
 r(x,t)=\bar{q}(\varepsilon_1x+x_0,\varepsilon_2t+t_0),                                                                \label{conjred}
\end{equation}
where $\varepsilon_j^2=1$, $j=1,2$, and $x_0, t_0\in \mathbb{R}$. Here the bar notation stands for
the complex conjugation.

We substitute the reduction \eqref{conjred} into \eqref{NLSq} and get
\begin{equation}
 iq_t(x,t)
 =q_{xx}(x,t)
 -2\sigma q^2(x,t) \bar{q}(\varepsilon_1x+x_0,\varepsilon_2t+t_0).        \label{firstconj}
\end{equation}
We then insert \eqref{conjred} into \eqref{NLSr}. We have
\begin{equation}
 i\frac{\partial}{\partial t}(\bar{q}(\varepsilon_1x+x_0,\varepsilon_2t+t_0))
 =-\frac{\partial^2}{\partial x^2}(\bar{q}(\varepsilon_1x+x_0,\varepsilon_2t+t_0))
 +2\sigma\bar{q}^2(\varepsilon_1x+x_0,\varepsilon_2t+t_0)q(x,t).
\end{equation}
Now introduce $\tilde{x}=\varepsilon_1x+x_0$, $\tilde{t}=\varepsilon_2t+t_0$ so that $x=\varepsilon_1(\tilde{x}-x_0)$, $t=\varepsilon_2(\tilde{t}-t_0)$
and take conjugation of the equation. This yields
\begin{equation}
 i\varepsilon_2q_{\tilde{t}}(\tilde{x},\tilde{t})
 =q_{\tilde{x}\tilde{x}}(\tilde{x},\tilde{t})
 -2\sigma q^2(\tilde{x},\tilde{t})\bar{q}(\varepsilon_1(\tilde{x}-x_0),\varepsilon_2(\tilde{t}-t_0)).            \label{secondred}
\end{equation}
To compare Eq. \eqref{secondred} with Eq. \eqref{firstconj} we write Eq. \eqref{firstconj} at $(\tilde{x},\tilde{t})$. Hence
these two equations are equivalent when
\begin{equation}
\varepsilon_2=1,\quad \varepsilon_1(\tilde{x}-x_0)=\varepsilon_1\tilde{x}+x_0,\quad \varepsilon_2(\tilde{t}-t_0)=\varepsilon_2\tilde{t}+t_0.
\end{equation}
yielding
\begin{equation}
 \varepsilon_2=1,\quad t_0=0,\quad (\varepsilon_1+1)x_0=0.          \label{conjconditions}
\end{equation}
Hence $x_0=0$ if $\varepsilon_1=1$, and it is free if $\varepsilon_1=-1$ for consistency. Thus, other than the local reduction ($(\varepsilon_1,\varepsilon_2)=(1,1)$),
the only consistent complex conjugate shifted nonlocal
reduction formula for the integrable NLS system  \eqref{NLSq}- \eqref{NLSr} is
\begin{equation}
 r(x,t)=\bar{q}(-x+x_0,t),\quad x_0\in \mathbb{R}.                   \label{validred}
\end{equation}
For completeness, we also check the shifted nonlocal reduction without complex conjugation
\begin{equation}
 r(x,t)=q(\varepsilon_1x+x_0,\varepsilon_2t+t_0).                               \label{realred}
\end{equation}
The NLS system \eqref{NLSq}-\eqref{NLSr} reduces consistently to shifted nonlocal equations if
$\varepsilon_2=-1$ and $\varepsilon_1=\pm 1$. If $\varepsilon_1=1$, then $x_0=0$. Hence we have two consistent shifted nonlocal reduction formulas in this case:
\begin{equation}
 r(x,t)=q(x,-t+t_0),\qquad
 r(x,t)=q(-x+x_0,-t+t_0).                         \label{validtime}
\end{equation}
Shifted nonlocal equations derived by \eqref{validred} and \eqref{validtime} are exactly the reverse space,
reverse time, and reverse space-time shifted nonlocal NLS reductions which have already been considered in
Refs.~\cite{AblowitzMusslimani2021}, \cite{AblowitzMusslimaniOssi2024}, \cite{GursesPekcan2022}. Since
these reductions consistently reduce the NLS system,
the resulting scalar shifted nonlocal equations inherit the corresponding integrable behavior.

In Eq.~\eqref{commented}, one has $\varepsilon_1=-1$ and
$\varepsilon_2=\lambda$ in the complex conjugate shifted nonlocal reduction \eqref{conjred}.  Hence,
\begin{itemize}
 \item[i.] For $\lambda=1$, to have a consistent reduction $t_0=0$.  The resulting
 equation is known as  reverse space shifted nonlocal NLS equation \cite{AblowitzMusslimaniOssi2024}, \cite{GursesPekcan2022}.
 Ref.~\cite{ZhouYu2026} treats $t_0$ as an arbitrary constant but for the integrable reduction of the standard integrable NLS system
 \eqref{NLSq}-\eqref{NLSr}, it is mandatory to have $t_0=0$.
 \item[ii.] For $\lambda=-1$, the reduction formula is a complex conjugate reverse space-time shifted nonlocal
 reduction. But it is not consistent with the standard integrable NLS system
 \eqref{NLSq}-\eqref{NLSr}. This does not make
 Eq.~\eqref{commented} ill-defined, but its integrability cannot be deduced
 from the standard integrable NLS system \eqref{NLSq}-\eqref{NLSr}. Ref.~\cite{ZhouYu2026} provides no valid proof or justification for the claimed integrability of Eq.~\eqref{commented} in the case $\lambda=-1$.
\end{itemize}

\section{Equivalence of shifted and unshifted equations for real shifts}

In this part we shall clarify that the real $x_0$ and $t_0$ shifts in the reflected variables can
be removed by a change of coordinates \cite{AblowitzMusslimaniOssi2024}. Let
\begin{equation}\label{coord}
 x=X+a,\qquad t=T+b,
\end{equation}
for some $a, b$ constants to be determined. Let also define
\begin{equation}
q(x,t)=Q(X,T).
\end{equation}
We simply have
\begin{equation}
Q(X,T)=q(X+a,T+b)\quad \mathrm{or}\quad q(x,t)=Q(x-a,t-b),
\end{equation}
due to \eqref{coord}. Then
\begin{align}
q(\varepsilon_1x+x_0,\varepsilon_2t+t_0)&=q(\varepsilon_1(X+a)+x_0,\varepsilon_2(T+b)+t_0)\nonumber\\
&=Q(\varepsilon_1(X+a)+x_0-a,\varepsilon_2(T+b)+t_0-b),
\end{align}
that is
\begin{align}
 \varepsilon_1x+x_0
 &\longmapsto \varepsilon_1X+x_0+(\varepsilon_1-1)a, \nonumber\\
 \varepsilon_2t+t_0
 &\longmapsto \varepsilon_2T+t_0+(\varepsilon_2-1)b.
\end{align}
Hence for $\varepsilon_1=-1$, choosing $a=\frac{x_0}{2}$; and similarly for $\varepsilon_2=-1$, taking $b=\frac{t_0}{2}$,  give
\begin{equation}
 -x+x_0\longmapsto -X,\quad -t+t_0\longmapsto -T.
\end{equation}
Thus, the three consistent shifted reductions become
\begin{equation}
\bar{q}(-x+x_0,t)= \bar{Q}(-X,T),\,\, q(x,-t+t_0)= Q(X,-T),\,\, q(-x+x_0,-t+t_0)= Q(-X,-T).
\end{equation}
Hence the real shift parameters $x_0$ and $t_0$ only determine the centers of the reflections.
They can be removed by the transformation $X=x-\frac{x_0}{2}$, $T=t-\frac{t_0}{2}$ and so they do not
produce new equations up to this transformation.
The relation between the shifted and unshifted equations and their solutions is considered in
Ref.~\cite{AblowitzMusslimaniOssi2024}.
Although a shifted equation is equivalent to its unshifted counterpart,  we observe the shifts
in the scattering data as phase factors if we study the shifted nonlocal equations in their original coordinates.

When $\lambda=-1$, by the transformation
\begin{equation}
 X=x-\frac{x_0}{2},\qquad T=t-\frac{t_0}{2},             \label{halfshift}
\end{equation}
the shifted nonlocal reverse space-time Eq.~\eqref{commented} becomes the following unshifted nonlocal reverse space-time equation
\begin{equation}
 iQ_T=Q_{XX}-2\sigma Q^2\bar{Q}(-X,-T).           \label{unshiftedlambda}
\end{equation}
Hence, the equation with $\lambda=-1$ can be reduced to its unshifted form
independent of whether it is integrable. The $\lambda=1$ case is
consistent only for $t_0=0$, while the remaining shift $x_0$ can be
eliminated through the change $X=x-\frac{x_0}{2}$.

\section{Clarification of the Hirota form}

In Hirota form for the equation \eqref{commented} given in Ref.~\cite{ZhouYu2026}
the starred quantities are written with the argument $(x,t)$, although the
subsequent calculations use them as shifted complex conjugates.  This can be
adjusted by defining
\begin{equation}
 \widehat h(x,t)=\bar{h}(-x+x_0,\lambda t+t_0).     \label{hatdef}
\end{equation}
Here the bar notation stands for complex conjugation. Hence, inserting the bilinearizing transformation
\begin{equation}
 q(x,t)=\frac{g(x,t)}{f(x,t)},\qquad \widehat q(x,t)=\frac{\widehat g(x,t)}{\widehat f(x,t)}=\frac{\bar{g}(-x+x_0,\lambda t+t_0)}{\bar{f}(-x+x_0,\lambda t+t_0)}                                                              \label{adj}
\end{equation}
into \eqref{commented} gives
\begin{equation}\displaystyle
i(g_tf-gf_t)-(g_{xx}f-2g_xf_x+gf_{xx})+\frac{2g}{f\widehat f}[\widehat f(ff_{xx}-f_x^2)+\sigma fg\widehat g]=0.
\end{equation}
The above equation yields the following Hirota form of Eq.~\eqref{commented} as
\begin{align}
 (iD_t-D_x^2)\{g \cdot f\}&=0,                    \label{bilinear1}\\
 \widehat f\,D_x^2\{f \cdot f\}
       +2\sigma f g\widehat g&=0.                     \label{bilinear2}
\end{align}
The bilinear
formulation is now algebraically consistent when the starred functions of
the commented article are viewed as the shifted conjugates of
Eq.~\eqref{hatdef}. It is important that the explicit arguments of those
functions be used; otherwise, under the direct interpretation of the star notation, the same expression
would be a bilinear form for the local NLS equation \cite{Hirota2004}.

To find $N$-soliton solutions of Eq.~\eqref{commented}, the authors of the paper \cite{ZhouYu2026} used the expansion
\begin{align}\label{expansion}
g=\sum_{n=1}^N\varepsilon^{2n-1}g_{2n-1},\,\, h=\sum_{n=1}^N\varepsilon^{2n-1}h_{2n-1},\,\, f=1+\sum_{n=1}^{N}\varepsilon^{2n}f_{2n}.
\end{align}
For one-soliton solution of Eq.~\eqref{commented}, inserting the above expansion into the Hirota form of the equation and analyzing the
coefficients of $\varepsilon^j$, $j=1, 2$ gave
\begin{equation}\label{gf_2}
g_1(x,t)=e^{kx-\omega t+\eta_0},\quad f_2=-\frac{\sigma}{(k-\bar{k})^2}e^{(k-\bar{k})x-(\omega+\lambda \bar{\omega})t+\bar{k}x_0-\bar{\omega}t_0+\eta_0+\bar{\eta}_0}
\end{equation}
with dispersion relation $\omega=ik^2$. After that point there is a significant problem in the application of the Hirota method. The coefficient of
$\varepsilon^3$ which is $(iD_t-D_x^2)\{g_1\cdot f_2\}$ does not vanish directly as stated in the commented paper \cite{ZhouYu2026}. Indeed using
(\ref{gf_2}) gives
\begin{equation}
\frac{\sigma e^{(2k-\bar{k})x+i(\lambda\bar{k}^2-2k^2)t+\bar{k}x_0+i\bar{k}^2t_0+2\eta_0+\bar{\eta}_0  }\bar{k}^2(1-\lambda)}{(k-\bar{k})^2}=0.
\end{equation}
Clearly, for $k\neq 0$ and $\sigma \neq 0$ the above relation is satisfied only for $\lambda=1$, but not for $\lambda=-1$. Consequently, one-soliton solution for the case $\lambda=-1$ derived in
\cite{ZhouYu2026} is not correct. For $\lambda=1$, the rest of the coefficients of $\varepsilon^j$ vanish directly. But still the solution for $\lambda=1$ has flaws.
Equation $(17)$ of \cite{ZhouYu2026} is neither derived from ordinary complex conjugation nor from the shifted complex conjugation. In the case when $f_2^*$ denotes $\bar{f}_2(x_0-x,t_0+t)$, the shifted complex conjugation has been wrongly done.

Another issue is that $\widehat q(x,t)=\bar{q}(-x+x_0,\lambda t+t_0)$ is treated as if it is a second
independent solution of the scalar shifted nonlocal equation, although it is completely
determined by the solution $q$ through the shifted nonlocal relation, that is
\begin{equation}
|\widehat q(x,t)| = |q(-x+x_0,\lambda t+t_0)|.
\end{equation}
Hence, the intensity graph of $\widehat q(x,t)$ can be derived from that of $q(x,t)$ simply by reflection and translation.
Thus, one of these surfaces cannot represent a localized wave while the
other one represents a periodic or non-localized wave as depicted in Figs.~1(a) and 1(b) of Ref.~\cite{ZhouYu2026}. The other inconsistency is about the value of $\sigma$. Although it is initially defined as $\sigma=\pm 1$, in the illustrations, different values like $\sigma=4$ and $\sigma=0.1$ are used in the commented paper.

\section{Conclusion}

In this work, we have examined the reduction, integrability, and Hirota-soliton claims made in Ref.~\cite{ZhouYu2026} for Eq.~\eqref{commented}.
The two values of $\lambda=\pm 1$ in Eq.~\eqref{commented} cannot both be justified as integrable reductions of the standard integrable NLS system
under the conditions stated in Ref.~\cite{ZhouYu2026}. For $\lambda=1$, consistency of the reduction requires $t_0=0$.
This mandatory criterion is not mentioned in the commented article \cite{ZhouYu2026}, but the parameter $t_0$ is improperly considered as an arbitrary constant. In case of $\lambda=-1$, the suggested criterion for the complex-conjugate reverse space-time shift is not consistent with the integrable NLS system and thus cannot be an integrable reduction of this system.

Inconsistency in the use of the Hirota's method is also observed. Namely, for $\lambda=-1$, the series expansion employed in \cite{ZhouYu2026} does not terminate as claimed: the coefficient of $\varepsilon^3$ yields a nonvanishing term for arbitrary wave number $k$. Hence the resulting formula for the one-soliton solution is not valid for $\lambda=-1$.
For the case $\lambda=1$, although the Hirota perturbation expansion works, the necessary condition $t_0=0$ is omitted and the expression for $f_2^*$ given in Eq. (17) of Ref.~\cite{ZhouYu2026} is incorrect.

Furthermore, the notation given in the Hirota form is misleading, the shifted-conjugate variables are not accurately presented, and the revised form has been provided in this work.

The function $\bar q(-x+x_0,\lambda t+t_0)$ is completely determined by $q(x,t)$ and is not an independent second solution. Thus, one of these two related surfaces cannot represent a localized wave while the other represents a periodic or non-localized wave, as depicted in Figs.~1(a) and 1(b) of Ref.~\cite{ZhouYu2026}. Another inconsistency is that, even though $\sigma$ is initially defined as $\sigma=\pm1$, in the figures, values such as $\sigma=4$ and $\sigma=0.1$ are used without any redefinition of its valid range.

We also clarified that all real shifts attached to reflected variables are removable by translations, so the
admissible shifted NLS equations are indeed equivalent to their unshifted
counterparts.

To sum up, these are significant mathematical contradictions in Ref.~\cite{ZhouYu2026}. No proper reduction, no Lax pair, no inverse scattering formulation, and no proper Hirota soliton construction have been given by \cite{ZhouYu2026} to show that Eq.~\eqref{commented} is integrable when $\lambda=-1$. Thus, the integrability claim is not  adequately supported for $\lambda=-1$ in \cite{ZhouYu2026}, and it still needs to be determined through an independent study of integrability.

\section{Acknowledgments}
  This work is partially supported by the Scientific
and Technological Research Council of Turkey (T\"{U}B\.{I}TAK).\\

\end{document}